\documentclass[a4paper,10pt,eqno]{article}
\usepackage{theorem}
\usepackage{latexsym,amssymb,amsfonts,amsmath}
\usepackage{graphicx}

\newtheorem{thm}{Theorem}[section]

\newtheorem{cor}[thm]{Corollary}

\newtheorem{pro}[thm]{Proposition}

\theoremheaderfont{\scshape}

\newcommand{\ZM}{\mathbb{Z}}

\title{{\Large {\bf STATIONARY MEASURE FOR TWO-STATE SPACE-INHOMOGENEOUS
QUANTUM WALK IN ONE DIMENSION}}}
\author{
{\small Hikari Kawai\footnote{kawai-hikari-dy@ynu.jp},\quad Takashi Komatsu\footnote{komatsu-takashi-fn@ynu.ac.jp (e-mail of the corresponding author)},\quad Norio Konno\footnote{konno-norio-bt@ynu.ac.jp}}\\
{\scriptsize  Department of Applied Mathematics, Faculty of Engineering, Yokohama National University}\\
{\scriptsize \footnotesize\it 79-5 Tokiwadai, Hodogaya, Yokohama, 240-8501, Japan}\\
}
\date{\empty}
\begin{document}
\maketitle

\par\noindent
\begin{small}
\par\noindent
{\bf Abstract}. We consider the two-state space-inhomogeneous coined quantum walk (QW) in one dimension. For a general setting, we obtain the stationary measure of the QW by solving the eigenvalue problem. As a corollary, stationary measures of the multi-defect model and space-homogeneous QW are derived. The former is a generalization of the previous works on one-defect model and the latter is a generalization of the result given by Konno and Takei (2015).

\footnote[0]{
{\it Abbr. title:} Stationary measures of three-state quantum walks on the one-dimensional lattice
}
\end{small}

\setcounter{equation}{0}

\section{Introduction \label{intro}}
The discrete time quantum walk (QW) was introduced as a quantum version of the classical random walk, whose time evolution are defined by unitary evolutions of amplitudes. The distribution of QW on the one dimensional lattice is different from that of the random walk \cite{Konno2002,Konno2005}. The review and books on QWs are Venegas-Andraca \cite{Venegas2013}, Konno \cite{Konno2008b}, Cantero et al. \cite{CanteroEtAl2013}, Portugal \cite{P2013}, Manouchehri and Wang \cite{MW2013}, for examples. The QW is a subject of study that has been investigated among quantum information and computation since around 2000. For its characteristic properties, recently, QWs have been widely studied by a number of groups in connection with various topics, for examples, separation of the radioisotope \cite{MatsuokaEtAl2011}, energy transfer of  photosynthesis complexes  \cite{MohseniEtAl2008} and topological insulator \cite{ObuseEtAl2015}.

There are two types of QWs, one is homogeneous QWs and the other is inhomogeneous QWs. The meaning of ``inhomogeneity" is that the quantum coin of a QW depends on time and/or space \cite{Ahl201214,Ahl201211,Ahl201152,joye2011}. We focus on space-inhomogeneous QWs in one dimension. One of the basic interests is to obtain measures induced by unitary evolutions of QWs, e.g., stationary measure, (time-averaged) limit measure and rescaled weak-limit measure. In this paper, we consider the stationary measures of QWs on $\mathbb{Z}$, where $\mathbb{Z}$ is the set of integers. Hence the stationary measure is the measure which does not depend on time. Especially, we get stationary measures of the two-state space-inhomogeneous QWs.
 
We briefly review the backgrounds of stationary measures for space-inhomogeneous models, i.e., QWs with defects. As for stationary measures of two-state QWs with one defect at the origin, Konno et al. \cite{KLS2013} showed that a stationary measure with exponential decay with respect to the position for the QW starting from infinite sites is identical to a time-averaged limit measure for the same QW starting from just the origin. We call this stationary measure a exponential type measure. One of our results contains the stationary measure shown by Konno et al. \cite{KLS2013}. Endo et al. \cite{ekst2014} got a stationary measure of the QW with one defect whose coin matrices are defined by the Hadamard matrix at $x \not= 0$ and the rotation matrix at $x=0$. Endo and Konno \cite{ek2014} calculated a stationary measure of the QW with one defect which was introduced and studied by W\'ojcik et al. \cite{WojcikEtAl2004}. Moreover, Endo et al. \cite{eko2015} and Endo et al. \cite{eekst2015} obtained stationary measures of the two-phase QW without defect and with one defect, respectively. Our result includes the stationary measure of the two-phase QW without defect and with one defect which was studied by Endo et al. \cite{eekst2015, eko2015}.

Konno and Takei \cite{kt2015} considered stationary measures of QWs and gave non-uniform stationary measures expressed as a quadratic polynomial. We call this stationary measure a quadratic polynomial type measure. Moreover, they proved that the set of the stationary measures contains uniform measure for the QW in general. So our aim is to find the non-trivial stationary measure of two-state QWs with multi-defect on $\mathbb{Z}$. One of our results belongs the  stationary measure with quadratic polynomial type which is given by Konno and Takei \cite{kt2015}.

Stationary measures for other QW models are also investigated, for example, three-state QW on $\mathbb{Z}$ \cite{EndoEtAl2016, EndoEtAl20162, Kawai2017, Konno2014, WangEtAl2015} and higher dimensional QW \cite{Komatsu2017}.  In order to analyze the details of QWs, a method based on transfer matrices is one of the common approaches, for example, Ahlbrecht et al. \cite{Ahl2011} and Bourget et al. \cite{Bou2003}. In this paper, we apply this method to two-state space-inhomogeneous QWs to obtain the stationary measures.

This paper is organized as follows. In Section \ref{Model}, we introduce the definition of the two-state inhomogeneous QWs with multi-defect on $\mathbb{Z}$. In Section \ref{result}, we present our results. Section \ref{proof} gives the proofs of results shown in the previous section by solving the corresponding the eigenvalue problem. In Section \ref{exam}, we deal with typical examples of two-state space-inhomogeneous QWs. Finally, summary is devoted to Section \ref{summ}.
\section{Model and method}\label{Model}
We introduce a discrete-time space-inhomogeneous QW on the line which is a quantum version of the classical random walk with an additional coin state. The particle has a coin state at time $n$ and position $x$ described by a two-dimensional vector:
\begin{align*}
\Psi_n (x)=
\begin{bmatrix}
\Psi_{n}^{L}(x)\\
\Psi_{n}^{R}(x)
\end{bmatrix}\ \ (x\in\mathbb{Z}),
\end{align*}
The upper and lower elements express left and right chiralities, respectively. The time evolution is determined by $2\times2$ unitary matrices $U_x$ which is called coin matrix here:
\begin{align*}
U_x=
\begin{bmatrix}
a_x&b_x\\
c_x&d_x
\end{bmatrix}
\ \ (x\in\mathbb{Z}).
\end{align*}
The subscript $x$ stands for the location. We divide $U_x$
into $U_x=P_x+Q_x$ with
\begin{align*}
P_x=
\begin{bmatrix}
a_x&b_x\\
0&0
\end{bmatrix},\ \ 
Q_x=
\begin{bmatrix}
0&0\\
c_x&d_x
\end{bmatrix}.
\end{align*}
The $2\times2$ matrix $P_x$ (resp. $Q_x$) represents that the walker moves to the left (resp. right) at position $x$ at each time step. Then the time evolution of the walk is defined by
\begin{align*}
\Psi_{n+1}(x)\equiv U^{(s)}\Psi_n(x)=P_{x+1}\Psi_n(x+1)+Q_{x-1}\Psi_n(x-1)\hspace{5mm}(x\in\mathbb{Z}).
\end{align*}
That is,
\begin{align*}
\begin{bmatrix}
\Psi_{n+1}^L (x)\\
\Psi_{n+1}^R (x)
\end{bmatrix}=
\begin{bmatrix}
a_{x+1}\Psi^L_n (x+1)+b_{x+1}\Psi^R_n (x+1)\\
c_{x-1}\Psi^L_n (x-1)+d_{x-1}\Psi^R_n (x-1)
\end{bmatrix}.
\end{align*}
Now let 
\begin{align*}
\Psi_n= {}^T[\cdots,\Psi_n^L (-1),\Psi_n^R (-1),\Psi_n^L (0),\Psi_n^R (0),
\Psi_n^L (1),\Psi_n^R (1),\cdots],
\end{align*}

\begin{align*}
U^{(s)}=
\begin{bmatrix}
\ddots&\vdots&\vdots&\vdots&\vdots&\vdots&\ldots\\
\ldots&O&P_{-1}&O&O&O&\ldots\\
\ldots&Q_{-2}&O&P_0&O&O&\ldots\\
\ldots&O&Q_{-1}&O&P_1&O&\ldots\\
\ldots&O&O&Q_0&O&P_2&\ldots\\
\ldots&O&O&O&Q_1&O&\ldots\\
\ldots&\vdots&\vdots&\vdots&\vdots&\vdots&\ddots
\end{bmatrix}
\ \ \ with\  O=
\begin{bmatrix}
0&0\\
0&0
\end{bmatrix},
\end{align*}
\noindent where $T$ means the transposed operation and the meaning
of the superscript $(s)$ is the first letter of system. Then the state of the QW at time $n$ is given by $\Psi_n =(U^{(s)})^n \Psi_0$ for any $n \geq 0$. Let $\mathbb{R}_{+} =[0,\infty).$ Here we introduce a map $\phi:(\mathbb{C}^2)^{\mathbb{Z}} \to \mathbb{R}^{\mathbb{Z}}_{+}$ such that for
\begin{align*}
\Psi={}^T\bigg[\cdots,
\begin{bmatrix}
\Psi^L (-1)\\
\Psi^R (-1)
\end{bmatrix},
\begin{bmatrix}
\Psi^L (0)\\
\Psi^R (0)
\end{bmatrix},
\begin{bmatrix}
\Psi^L (1)\\
\Psi^R (1)
\end{bmatrix},
\cdots\bigg]\in(\mathbb{C}^2)^{\mathbb{Z}},
\end{align*}
we define the measure of the QW by 
$
\mu:\mathbb{Z} \to \mathbb{R}_{+}
$
satisfying
\begin{align*}  
\mu(x)=\phi(\Psi)(x)=|\Psi^L(x)|^2+|\Psi^R(x)|^2 \ \ \ (x\in \mathbb{Z}).
\end{align*}
We should note that $\mu(x)$ gives the measure of the QW at position $x$. Our model here is considered on the set of all the $\mathbb{C}^2$-valued functions on $\mathbb{Z}$ whose inner product is $\langle\Psi,\Phi\rangle =\sum_{x\in\mathbb{Z}}\langle\Psi(x),\Phi(x)\rangle_{\mathbb{C}^2}$, where $\langle\cdot,\cdot\rangle_{\mathbb{C}^2}$ denotes the standard inner product on $\mathbb{C}^2$. We do not have any assumptions on the norm for the sets. In this paper, we consider the stationary measures for QWs on the above framework. 

Let $\cal{M}$$(U^{(s)})$ be the set of measures of the QW. To explain our results, we introduce three classes of the measures for QW. First one is the set of the measures with exponential type:
\begin{equation*}
\begin{split}
&\mathcal{M}_{et}(U^{(s)})=\Big\{\mu\in\mathcal{M}(U^{(s)})\ ;\ there\ exist\ c_{+},\ c_{-}>0\ (c_{+}, c_{-}\ne1)\ such\ that\\
&\hspace{4.0cm}0<\lim_{x\to+\infty}\frac{\mu(x)}{c_{+}^x}<+\infty,\quad 0<\lim_{x\to-\infty}\frac{\mu(x)}{c_{-}^x}<+\infty\Big\},
\end{split}
\end{equation*}
where $\mathcal{M}(U^{(s)})$ is the set of measures on $\mathbb{Z}$. Second one is the set of the measures with quadratic polynomial type:
\begin{equation*}
\begin{split}
&\mathcal{M}_{qpt}(U^{(s)})=\Big\{\mu\in\mathcal{M}(U^{(s)})\ ;\ 0<\lim_{x\to\pm\infty}\frac{\mu(x)}{|x|^2}<+\infty\Big\}.
\end{split}
\end{equation*}
Last one is the set of the stationary measures:
\begin{equation*}
\begin{split}
&\mathcal{M}_s(U^{(s)})=\Big\{\mu\in\mathcal{M}(U^{(s)})\ ;\ there\ exists\ \Psi_0\in\left(\mathbb{C}^2\right)^{\mathbb{Z}}\ such\ that\\
&\hspace{6.0cm}\phi{((U^{(s)}})^n\Psi_0)=\mu\ (n=0,1,2,\ldots)\Big\}
\end{split}
\end{equation*}
and we call the element of ${\cal{M}}_s(U^{(s)})$ the stationary measure of the QW. In general, if unitary operators $U^{(s)}_1$ and $U^{(s)}_2$ are different, the sets of stationary measures $\mathcal{M}_s(U^{(s)}_1)$ and $\mathcal{M}_s(U^{(s)}_2)$ are different. For example, if we take the unitary operators $U^{(s)}_1$ and $U^{(s)}_2$ corresponding to the following matrices $U_1$ and $U_2$ respectively:
\begin{align*}
U_1=
\begin{bmatrix}
1&0\\
0&1
\end{bmatrix}, \ \ \ 
U_2=\frac{1}{\sqrt{2}}
\begin{bmatrix}
1&1\\
1&-1
\end{bmatrix},
\end{align*}
then we have
\begin{align*}
\mathcal{M}_s(U^{(s)}_1)=\mathcal{M}_{unif}(U^{(s)}),\ \ \ 
\mathcal{M}_s(U^{(s)}_2)\supsetneq\mathcal{M}_{unif}(U^{(s)}).
\end{align*}
The above results are given in Konno and Takei \cite{kt2015}. Here $\mathcal{M}_{unif}(U^{(s)})$ is the set of the uniform measures defined by
\[
\mathcal{M}_{unif}(U^{(s)})=\Big\{\mu_{c}\in\mathcal{M}(U^{(s)})\ ;\ \mu_{c}(x)=c,\ c>0 \Big\}.
\]
Let us consider the eigenvalue problem:
\begin{align*}
U^{(s)}\Psi=\lambda\Psi,
\end{align*}
where $\lambda\in\mathbb{C}$ with $|\lambda|=1$ and $U^{(s)}$ is an doubly infinite unitary matrix. If we assume that the initial state $\Psi_0$ is the above solution, we have 
\begin{align*}
\Psi_n =(U^{(s)})^n \Psi_0 =\lambda^n \Psi_0.
\end{align*}
Noting that $|\lambda|=1$, we see
\begin{align*}
\mu_n (x) =||\Psi_n (x)||^2=|\lambda|^{2n}||\Psi_0 (x)||^2 =\mu_0 (x)\ \ (x\in\mathbb{Z}).
\end{align*}
Therefore $\mu_0 (x)=\phi(\Psi_0)(x)$ gives the stationary measure.


\section{Results}\label{result}
Applying the method introduced in Section 4 to the space-inhomogeneous QW, we solve the eigenvalue problem $U^{(s)}\Psi=\lambda\Psi$ as follows. From now on, we put $\alpha=\Psi^L (0)$ and $\beta=\Psi^R (0)$.

\begin{thm}\label{thm:3.1}
Let $\Psi(x)={}^T[\Psi^L (x),\Psi^R (x)]$ be the amplitude.
Put coin matrix which is defined by
\begin{align*}
U_x=
\begin{bmatrix}
a_x&b_x\\
c_x&d_x
\end{bmatrix}\ \ (x\in\mathbb{Z}),
\end{align*}
where $a_x b_x c_x d_x\neq 0$. Then a solution of the following eigenvalue problem:
\begin{align*}
U^{(s)}\Psi=\lambda \Psi
\end{align*}
is given by
\begin{eqnarray*}
\Psi(x)=
\begin{cases}
\displaystyle \prod_{y=1}^{x} D^{+}_{y} \Psi(0)&(x\geq 1),\\
\Psi(0)&(x=0),\\
\displaystyle \prod_{y=-1}^{x} D^{-}_{y} \Psi(0)&(x\leq -1),
\end{cases}
\end{eqnarray*}
where
\begin{eqnarray*}
D^{+}_{x}=\begin{bmatrix}
\dfrac{\lambda^{2}-b_x c_{x-1}}{\lambda a_x}&-\dfrac{b_x d_{x-1}}{\lambda a_x}\\
\dfrac{c_{x-1}}{\lambda}&\dfrac{d_{x-1}}{\lambda}
\end{bmatrix},\ \ 
D^{-}_{x}=\begin{bmatrix}
\dfrac{a_{x+1}}{\lambda}&\dfrac{b_{x+1}}{\lambda}\\
-\dfrac{a_{x+1}c_x}{\lambda d_x}&\dfrac{\lambda^2-b_{x+1} c_{x}}{\lambda d_x}
\end{bmatrix}.
\end{eqnarray*}
Moreover a stationary measure $\mu$ is given by
\begin{align*}
\mu (x)=\phi(\Psi)(x)=||\Psi(x)||^2\ \ \ (x\in\mathbb{Z}).
\end{align*}
\end{thm}

We introduce some notations
\begin{eqnarray*}
&\mathbb{Z}_{\geq}=\{0,1,2,\ldots \},&\\
&\mathbb{Z}_{>}=\{1,2,3,\ldots \},&\\
&\mathbb{Z}_{[a,b]}=\{a, a+1, \ldots, b-1, b\}&(a,b\in\mathbb{Z}\ with\ a < b).
\end{eqnarray*}  
Next we consider a special case of Theorem \ref{thm:3.1} in which a sequence of coin matrices $\{U_x\}$ is defined by
\begin{align*}
U_x=U\ \ (x\notin\mathbb{Z}_{[-m,n]})
\end{align*}
for $m,n\in\mathbb{Z}_>$. Here $U$ is a $2\times2$ unitary matrix. The model can be considered on a QW with $(m+n+1)$ defects. The following result is a direct consequence of Theorem 3.1.
\begin{pro}\label{pro:3.2}
Put $m,n\in\mathbb{Z}_{>}$.
Let $\Psi(x)={}^T[\Psi^L (x),\Psi^R (x)]$ be the amplitude.
Define a sequence of coin matrices $\{U_x\}$ as follows:
\begin{align*}
&U_x=
\begin{cases}
\begin{bmatrix}
a_x&b_x\\
c_x&d_x
\end{bmatrix}&(x\in\mathbb{Z}_{[-m,n]}),\vspace{3mm}\\
\begin{bmatrix}
a&b\\
c&d
\end{bmatrix}&(x\notin\mathbb{Z}_{[-m,n]}),
\end{cases}
\end{align*}
where $a_x b_x c_x d_x\neq0 $, and $abc d\neq0$. Then a solution of the following eigenvalue problem:
\begin{align*}
U^{(s)}\Psi=\lambda \Psi
\end{align*}
is given by
\begin{align*}
\Psi(x)=
\begin{cases}
\bigg\{ (D^{+}) ^{x-(n+1)}\bigg\}\displaystyle \prod_{y=1}^{n+1}D^{+}_{y} \Psi(0)&(n+2\leq x),\\
\displaystyle \prod_{y=1}^{x}D^{+}_{y} \Psi(0)&(1\leq x \leq n+1),\\
\Psi(0)&(x=0),\\
\displaystyle \prod_{y=-1}^{x}D^{-}_{y} \Psi(0)&(-(m+1) \leq x \leq -1),\\
\bigg\{ (D^{-})^{-x-(m+1)} \bigg\}\displaystyle \prod_{y=-1}^{-(m+1)}D^{-}_{y} \Psi(0)&(x\leq -(m+2)),
\end{cases}
\end{align*}
where
\begin{eqnarray*}
D^{+}=
\begin{bmatrix}
\dfrac{\lambda^2-bc}{\lambda a}&-\dfrac{bd}{\lambda a}\\
\dfrac{c}{\lambda}&\dfrac{d}{\lambda}
\end{bmatrix},\ 
D^{+}_{n+1}=
\begin{bmatrix}
\dfrac{\lambda^{2}-b c_{n}}{\lambda a}&-\dfrac{b d_{n}}{\lambda a}\\
\dfrac{c_{n}}{\lambda}&\dfrac{d_{n}}{\lambda}
\end{bmatrix},\\\\
D^{-}_{-(m+1)}=
\begin{bmatrix}
\dfrac{a_{-m}}{\lambda}&\dfrac{b_{-m}}{\lambda}\\
-\dfrac{a_{-m} c}{\lambda d}&\dfrac{\lambda^2-b_{-m} c}{\lambda d}
\end{bmatrix},\ 
D^{-}=
\begin{bmatrix}
\dfrac{a}{\lambda}&\dfrac{b}{\lambda}\\
-\dfrac{ac}{\lambda d}&\dfrac{\lambda^2-bc}{\lambda d}
\end{bmatrix}.
\end{eqnarray*}
Furthermore, a stationary measure $\mu$ is determined by
\begin{align*}
\mu (x)=\phi(\Psi)(x)=||\Psi(x)||^2\ \ \ (x\in\mathbb{Z}).
\end{align*}
\end{pro}

The following result is obtained by solving the recurrence relations of $D^+$ and $D^-$, respectively.
\begin{pro}\label{pro:3.3}
We put $\Psi^{L}(x)=S_x, \Psi^{R}(x)=T_x\ and\ \Delta=ad-bc$.
\begin{description}
\item[(1)]
For $x\geq n+1$, we have
\begin{description}
\item[(i)] $\lambda^2 \neq ad+bc\pm2\sqrt{abcd}$ case
\begin{align*}
S_x&=\dfrac{1}{\Lambda_{+} -\Lambda_{-}}\Big\{
\Lambda^{x-n}_{+} ( S_{n+1}-\Lambda_{-} S_{n} ) -
\Lambda^{x-n}_{-} ( S_{n+1}-\Lambda_{+} S_{n} )\Big\},&\\\\
T_{x}&
=\dfrac{1}{\Lambda_{+} -\Lambda_{-}}\Big\{
\Lambda^{x-(n+1)}_{+} ( T_{n+2}-\Lambda_{-} T_{n+1} ) -\\
&\hspace{50mm}\Lambda^{x-(n+1)}_{-} ( T_{n+2}-\Lambda_{+} T_{n+1} )\Big\},
\end{align*}
where
\begin{align*}
\Lambda_{\pm}=\dfrac{\lambda^2+\Delta\pm\sqrt{(\lambda^2+\Delta)^2-4\lambda^{2}ad}}{2a\lambda}.
\end{align*}
\item[(ii)] $\lambda^2 = ad+bc\pm2\sqrt{abcd}$ case
\begin{align*}
S_x &= \Lambda^{x-(n+1)} \Big[S_{n+1}+\{x-(n+1)\}{(S_{n+1}-\Lambda S_n)}\Big]\vspace{2mm},&\\
T_x &= \Lambda^{x-(n+2)} \Big[T_{n+2}+\{x-(n+2)\}{(T_{n+2}-\Lambda T_{n+1})}\Big],&
\end{align*}
where\begin{align*}
\Lambda=\dfrac{\lambda^2+\Delta}{2a\lambda}.
\end{align*}
\end{description}
\item[(2)]
For $x\leq -(m+1)$, we have
\begin{description}
\item[(i)] $\lambda^2 \neq ad+bc\pm2\sqrt{abcd}$ case
\begin{align*}
S_{x}&=\dfrac{1}{\Gamma_{+}-\Gamma_{-}}\bigg\{
\Gamma^{-x-(m+1)}_{+} ( S_{-(m+2)}-\Gamma_{-} S_{-(m+1)} )\\
&\ \ \ \ \ \ \ \ \ \ \ \ \ \hspace{22mm}-
\Gamma^{-x-(m+1)}_{-} ( S_{-(m+2)}-\Gamma_{+} S_{-(m+1)} )\bigg\},
\\\\
T_{x}&=\dfrac{1}{\Gamma_{+}-\Gamma_{-}}
\bigg\{ \Gamma^{-x-m}_{+} ( T_{-(m+1)}-\Gamma_{-} T_{-m} )\\
&\hspace{50mm}-
\Gamma^{-x-m}_{-} ( T_{-(m+1)}-\Gamma_{+} T_{-m} )\bigg\},
\end{align*}
where
\begin{align*}
\Gamma_{\pm}=\dfrac{\lambda^2+\Delta\pm\sqrt{(\lambda^2+\Delta)^2-4\lambda^{2}ad}}{2d\lambda}=\dfrac{d}{a} \Lambda_{\pm}.
\end{align*}
\item[(ii)] $\lambda^2 = ad+bc\pm2\sqrt{abcd}$ case
\begin{align*}
S_x &=\Gamma^{-x-(m+2)} \Big[ S_{-(m+2)}+\{-x-(m+2)\} (S_{-(m+2)}-\Gamma S_{-(m+1)})\Big]\vspace{2mm},&\\
T_x &= \Gamma^{-x-(m+1)} \Big[T_{-(m+1)}+\{-x-(m+1)\} (T_{-(m+1)}-\Gamma T_{-m})\Big],&
\end{align*}
where
\begin{align*}
\Gamma=\dfrac{\lambda^2+\Delta}{2d\lambda}.
\end{align*}
\end{description}
\end{description}
\end{pro}

From this theorem we can obtain the following result for the space-homogeneous case. 
\begin{cor}\label{cor:3.4}
Let $\Psi(x)={}^T[\Psi^L (x),\Psi^R (x)]$ be the amplitude.
Put
\begin{align*}
U_x=U=
\begin{bmatrix}
a&b\\
c&d
\end{bmatrix}\ \ (x\in\mathbb{Z}),
\end{align*}
with $abcd\neq0$. Then a solution of the following eigenvalue problem:
\begin{align*}
U^{(s)}\Psi=\lambda \Psi
\end{align*}
is given by
\begin{description}
\item[(i)] $\lambda^2 \neq ad+bc\pm2\sqrt{abcd}$ case
\begin{align*}
&\!\!\!\!\!\begin{bmatrix}
\Psi^{L}(x)\\
\Psi^{R}(x)
\end{bmatrix}\\
&=
\begin{cases}
\dfrac{1}{\Lambda_{+} -\Lambda_{-}}
\begin{bmatrix}
\Lambda^{x}_{+} (\Psi^{L}(1)-\Lambda_{-}\alpha)-
\Lambda^{x}_{-}(\Psi^{L}(1)-\Lambda_{+}\alpha)\vspace{3mm}\\
\Lambda^{x}_{+} (\Psi^{R}(1)-\Lambda_{-}\beta)-
\Lambda^{x}_{-}(\Psi^{R}(1)-\Lambda_{+}\beta)
\end{bmatrix}&(x\geq1),\\\\
\dfrac{1}{\Gamma_{+} -\Gamma_{-}}
\begin{bmatrix}
\Gamma^{-x}_{+} (\Psi^{L}(-1)-\Gamma_{-}\alpha)-
\Gamma^{-x}_{-}(\Psi^{L}(-1)-\Gamma_{+}\alpha)\vspace{3mm}\\
\Gamma^{-x}_{+} (\Psi^{R}(-1)-\Gamma_{-}\beta)-
\Gamma^{-x}_{-}(\Psi^{R}(-1)-\Gamma_{+}\beta)
\end{bmatrix}&(x\leq-1),
\end{cases}
\end{align*}\\
\item[(ii)]$\lambda^2 = ad+bc\pm2\sqrt{abcd}$ case
\begin{align*}
&\!\!\!\!\!\begin{bmatrix}
\Psi^{L}(x)\\
\Psi^{R}(x)
\end{bmatrix}\\
&=
\begin{cases}
\bigg(\dfrac{\lambda^2+\Delta}{2a\lambda}\bigg)^x \dfrac{1}{\lambda^2+\Delta}
\begin{bmatrix}
\alpha(1+x)\lambda^2-(\alpha\nabla+2bd\beta)x+\alpha\Delta\vspace{3mm}\\
\beta(1-x)\lambda^2+(\beta\nabla+2ac\alpha)x+\beta\Delta
\end{bmatrix}&(x\geq1),\\\\
\bigg(\dfrac{\lambda^2+\Delta}{2d\lambda}\bigg)^{-x} \dfrac{1}{\lambda^2+\Delta}
\begin{bmatrix}
\alpha(1+x)\lambda^2-(\alpha\nabla+2bd\beta)x+\alpha\Delta\vspace{3mm}\\
\beta(1-x)\lambda^2+(\beta\nabla+2ac\alpha)x+\beta\Delta
\end{bmatrix}&(x\leq-1),
\end{cases}
\end{align*}
\end{description}
where
\begin{align*}
&\Delta=ad-bc,\ 
\nabla=ad+bc,&\\
&\Lambda_{\pm}=\dfrac{\lambda^2+\Delta\pm\sqrt{(\lambda^2+\Delta)^2-4\lambda^{2}ad}}{2a\lambda},\ 
\Gamma_{\pm}=\dfrac{\lambda^2+\Delta\pm\sqrt{(\lambda^2+\Delta)^2-4\lambda^{2}ad}}{2d\lambda},&\\
&\Psi^{L}(1)=\dfrac{\alpha \lambda^2-b(d\beta+c\alpha)}{a\lambda},\ 
\Psi^{R}(1)=\dfrac{d\beta+c\alpha}{\lambda},&\\
&\Psi^{L}(-1)=\dfrac{b\beta+a\alpha}{\lambda},\ 
\Psi^{R}(-1)=\dfrac{\beta \lambda^2-c(b\beta+a\alpha)}{d\lambda}.&
\end{align*}
Furthermore, a stationary measure $\mu$ is given by
\begin{align*}
\mu (x)=\phi(\Psi)(x)=||\Psi(x)||^2\ \ \ (x\in\mathbb{Z}).
\end{align*}
\end{cor}
\noindent A stationary measure $\mu\in\mathcal{M}(U^{(s)})$ in case (i) of Corollary \ref{cor:3.4} becomes an exponential type measure, i.e., $\mu\in\mathcal{M}_s(U^{(s)})\cap\mathcal{M}_{et}(U^{(s)})$. On the other hand, a stationary measure $\mu\in\mathcal{M}(U^{(s)})$ in case (ii) of Corollary \ref{cor:3.4} becomes a quadratic polynomial type measure, i.e. $\mu\in\mathcal{M}_s(U^{(s)})\cap\mathcal{M}_{qpt}(U^{(s)})$, which was given in Konno and Takei \cite{kt2015}.

\section{Proofs}\label{proof}
In this section, we prove Theorem \ref{thm:3.1} and Proposition \ref{pro:3.3}.
\subsection{Proof of Theorem \ref{thm:3.1}}
We focus on the space-inhomogeneous QW whose coin matrix is determined by
\begin{align*}
U_x=
\begin{bmatrix}
a_x&b_x\\
c_x&d_x
\end{bmatrix}&\ \ (x\in\mathbb{Z})
\end{align*}
where $a_x b_x c_x d_x\neq 0 $. We consider the solution of
\begin{align*}
U^{(s)}\Psi=\lambda\Psi.
\end{align*}
Then $\Psi(x)={}^T[\Psi^L (x),\Psi^R (x)]$ satisfies
 \begin{align}
\lambda
\begin{bmatrix}
\Psi^{L}(x)\\
\Psi^{R}(x)
\end{bmatrix}=
\begin{bmatrix}
a_{x+1}\Psi^{L}(x+1)+b_{x+1}\Psi^{R}(x+1)\\
c_{x-1}\Psi^{L}(x-1)+d_{x-1}\Psi^{R}(x-1)
\end{bmatrix}\ \ (x\in\mathbb{Z})\label{nagi1}.
\end{align}
From now on, we will obtain $D^+$ and $D^-$. Eq.(\ref{nagi1}) gives
\begin{align}
\Psi^L(x-1)=\dfrac{a_{x}}{\lambda}\Psi^L(x)+\dfrac{b_x}{\lambda}\Psi^R(x).\label{nagi2}
\end{align}
Then Eq.(\ref{nagi2}) can be rewritten as
\begin{align*}
\Psi^L(x-1)=\dfrac{a_x}{\lambda}\Psi^L(x)+\dfrac{b_x}{\lambda}
\bigg\{ \dfrac{c_{x-1}}{\lambda}\Psi^L(x-1)+\dfrac{d_{x-1}}{\lambda}\Psi^R(x-1) \bigg\}.
\end{align*}
Hence, we get
\begin{align*}
\Psi^L(x)=\dfrac{1}{a_x}\left( \lambda-\dfrac{b_x c_{x-1}}{\lambda}\right) \Psi^L(x-1)
-\dfrac{b_x d_{x-1}}{a_x \lambda}\Psi^R(x-1).
\end{align*}
From now on, we put
\begin{align*}
D^+_x=
\begin{bmatrix}
\dfrac{\lambda^2-{b_x c_{x-1}}}{a_x \lambda}&-\dfrac{b_x d_{x-1}}{a_x \lambda}\\
\dfrac{c_{x-1}}{\lambda}&\dfrac{d_{x-1}}{\lambda}
\end{bmatrix}.
\end{align*}
Therefore Eq.(\ref{nagi1}) becomes
\begin{align*}
\begin{bmatrix}
\Psi^L(x)\\
\Psi^R(x)
\end{bmatrix}&=
\begin{bmatrix}
\dfrac{1}{a_x}\left(\lambda-\dfrac{b_x c_{x-1}}{\lambda}\right)\Psi^L(x-1)
-\dfrac{b_x d_{x-1}}{a_x \lambda}\Psi^R(x-1)\\
\dfrac{c_{x-1}}{\lambda}\Psi^{L}(x-1)+\dfrac{d_{x-1}}{\lambda}\Psi^{R}(x-1)
\end{bmatrix}&
\\&=\begin{bmatrix}
\dfrac{\lambda^2-{b_x c_{x-1}}}{a_x \lambda}&-\dfrac{b_x d_{x-1}}{a_x \lambda}\\
\dfrac{c_{x-1}}{\lambda}&\dfrac{d_{x-1}}{\lambda}
\end{bmatrix}
\begin{bmatrix}
\Psi^L(x-1)\\
\Psi^R(x-1)
\end{bmatrix}&\\
&=D^+_x
\begin{bmatrix}
\Psi^L(x-1)\\
\Psi^R(x-1)
\end{bmatrix}.&
\end{align*}
Thus we get
\begin{align}
\Psi(x)=D^+_x\Psi(x-1)\ \ \ (x\in\mathbb{Z})\label{nagi3}.
\end{align}
From Eq.(\ref{nagi3}), we obtain
\begin{align*}
\Psi(x)=\displaystyle \prod_{y=1}^{x}D^+_y \Psi(0)\ \ \ (x\geq1).
\end{align*}
We put $\{D^+_x\}^{-1}=D^-_{x-1}$. We should remark that determinant of $D^{+}_{x}$ is not $0$, since $a_x\neq 0\ (x\in\mathbb{Z})$. Then Eq.(\ref{nagi3}) gives
\begin{align}
\Psi(x)=D^-_{x} \Psi(x+1)\ \ \ (x\in\mathbb{Z}).\label{nagi4}
\end{align}
By Eq.(\ref{nagi4}), we have
\begin{align*}
\Psi(x)=\displaystyle \prod_{y=-1}^{x}D^-_y \Psi(0)\ \ \ (x\leq-1).
\end{align*}

\subsection{Proof of Proposition 3.3}
From now on, we focus on the space-inhomogeneous QW whose coin matrix is determined by
\begin{align*}
U_x=
\begin{cases}
\begin{bmatrix}
a_x&b_x\\
c_x&d_x
\end{bmatrix}&(x\in\mathbb{Z}_{[-m,n]}),\vspace{3mm}\\
\begin{bmatrix}
a&b\\
c&d
\end{bmatrix}&(x\notin\mathbb{Z}_{[-m,n]}),
\end{cases}
\end{align*}
where $a_x b_x c_x d_x\neq 0 ,$ and $ abcd\neq0$. We consider the solution of
\begin{align*}
U^{(s)}\Psi=\lambda\Psi.
\end{align*}
Then $\Psi(x)={}^T[\Psi^L (x),\Psi^R (x)]$ satisfies
\begin{align*}
\lambda
\begin{bmatrix}
\Psi^L (x)\\
\Psi^R (x)
\end{bmatrix}=
\begin{bmatrix}
0&0\\
c_{x-1}&d_{x-1}
\end{bmatrix}
\begin{bmatrix}
\Psi^L (x-1)\\
\Psi^R (x-1)
\end{bmatrix}+
\begin{bmatrix}
a_{x+1}&b_{x+1}\\
0&0
\end{bmatrix}
\begin{bmatrix}
\Psi^L (x+1)\\
\Psi^R (x+1)
\end{bmatrix}\ \ \ (x\in\mathbb{Z}).
\end{align*}
First we consider $x\geq0$ case. In particular, we have
\begin{align*}
\lambda
\begin{bmatrix}
\Psi^L (x)\\
\Psi^R (x)
\end{bmatrix}=
\begin{bmatrix}
0&0\\
c&d
\end{bmatrix}
\begin{bmatrix}
\Psi^L (x-1)\\
\Psi^R (x-1)
\end{bmatrix}+
\begin{bmatrix}
a&b\\
0&0
\end{bmatrix}
\begin{bmatrix}
\Psi^L (x+1)\\
\Psi^R (x+1)
\end{bmatrix}\hspace{5mm}(x\geq n+2),
\end{align*}
From now on, we put  $S_x =\Psi^L (x)$ and $T_x=\Psi^R (x)$, then above equations become
\begin{eqnarray}
&\lambda S_x=a S_{x+1}+	b T_{x+1}&(x\geq n)\label{ukiyoe1},\\
&\lambda T_x=c S_{x-1}+d T_{x-1}&(x\geq n+2)\label{ukiyoe2}.
\end{eqnarray}
Then Eq.(\ref{ukiyoe1}) and Eq.(\ref{ukiyoe2}) can be rewritten as
\begin{eqnarray}
&\lambda S_{x-1}=a S_x+b T_x &(x\geq n+1)\label{ukiyoe3},\\
&T_x=\dfrac{c}{\lambda}S_{x-1} +\dfrac{d}{\lambda}T_{x-1}&(x\geq n+2)\label{ukiyoe4}.
\end{eqnarray}
From Eq.(\ref{ukiyoe3}), we have
\begin{eqnarray}
&T_x=\dfrac{\lambda}{b} S_{x-1}-\dfrac{a}{b}S_x &(x\geq n+1)\label{ukiyoe5},\\
&T_{x-1}=\dfrac{\lambda}{b}S_{x-2}-\dfrac{a}{b}S_{x-1} &(x\geq n+2)\label{ukiyoe6}.
\end{eqnarray}
By substituting Eqs. (\ref{ukiyoe5}) and (\ref{ukiyoe6}) into Eq. (\ref{ukiyoe3}), we see that $S_x$ satisfies
\begin{eqnarray}
\lambda \dfrac{a}{b} S_x+\left( c-\dfrac{ad}{b}-\dfrac{\lambda^2}{b} \right) S_{x-1} +\lambda\dfrac{ d}{b}S_{x-2} =0\hspace{5mm}(x\geq n+2)\label{ukiyoe7}.
\end{eqnarray} 
If the characteristic equation of Eq.(\ref{ukiyoe7}) has two distinct roots, $\Lambda_+$ and $\Lambda_-$, then we obtain
\begin{eqnarray}
\begin{cases}
S_x-\Lambda_+ S_{x-1} =\Lambda_- (S_{x-1} - \Lambda_+ S_{x-2})\\
S_x-\Lambda_- S_{x-1} =\Lambda_+ (S_{x-1} - \Lambda_- S_{x-2})
\end{cases}\hspace{5mm}(x\geq n+1)\label{ukiyoe8},
\end{eqnarray}
where
\begin{align*}
\Lambda_{\pm}=\dfrac{\lambda^2+\Delta\pm\sqrt{(\lambda^2+\Delta)^2-4\lambda^{2}ad}}{2a\lambda},
\end{align*}
with
\begin{align*}
\Delta=ad-bc.
\end{align*}
Then Eq.(\ref{ukiyoe8}) can be rewritten as
\begin{eqnarray*}
\begin{cases}
S_{x+1}-\Lambda_+ S_x =\Lambda^{x-n}_- (S_{n+1}-\Lambda_+ S_n)\\
S_{x+1}-\Lambda_- S_x =\Lambda^{x-n}_+ (S_{n+1}-\Lambda_- S_n)
\end{cases}\hspace{5mm}(x\geq n+1),
\end{eqnarray*}
hence
\begin{eqnarray*}
S_x&=\dfrac{1}{\Lambda_{+} -\Lambda_{-}}\Big\{
\Lambda^{x-n}_{+} ( S_{n+1}-\Lambda_{-} S_{n} ) -
\Lambda^{x-n}_{-} ( S_{n+1}-\Lambda_{+} S_{n} )\Big\}\hspace{5mm}(x\geq n).&
\end{eqnarray*}
Next, if the characteristic equation of Eq.(\ref{ukiyoe7}) has a multiple root, $\Lambda$,  then we have
\begin{eqnarray}
S_x -\Lambda S_{x-1}=\Lambda (S_{x-1}-\Lambda S_{x-2})\hspace{5mm}(x\geq n+2)\label{DRI1},
\end{eqnarray}
where
\begin{align*}
\Lambda=\dfrac{\lambda^2+\Delta}{2a\lambda}.
\end{align*}
Then Eq.(\ref{DRI1}) implies
\begin{eqnarray}
S_x -\Lambda S_{x-1} =\Lambda^{x-(n+1)} (S_{n+1}-\Lambda S_n)\hspace{5mm}(x\geq n+2)\label{DRI2}.
\end{eqnarray}
From Eq.(\ref{DRI2}), we get
\begin{eqnarray}
S_x = \Lambda^{x-(n+1)} \Big[S_{n+1}+\{x-(n+1)\}{(S_{n+1}-\Lambda S_n)}\Big]\hspace{5mm}(x\geq n).
\end{eqnarray}
Therefore 
\begin{description}
\item[(i)] $\lambda^2 \neq ad+bc\pm2\sqrt{abcd}$ case
\begin{align*}
S_x&=\dfrac{1}{\Lambda_{+} -\Lambda_{-}}\Big\{
\Lambda^{x-n}_{+} ( S_{n+1}-\Lambda_{-} S_{n} ) -
\Lambda^{x-n}_{-} ( S_{n+1}-\Lambda_{+} S_{n} )\Big\}\hspace{5mm}(x\geq n)\label{ukiyoe9},&
\end{align*}
where
\begin{align*}
\Lambda_{\pm}=\dfrac{\lambda^2+\Delta\pm\sqrt{(\lambda^2+\Delta)^2-4\lambda^{2}ad}}{2a\lambda}.
\end{align*}
\item[(ii)] $\lambda^2 = ad+bc\pm2\sqrt{abcd}$ case
\begin{align*}
S_x &= \Lambda^{x-(n+1)} \Big[S_{n+1}+\{x-(n+1)\}{(S_{n+1}-\Lambda S_n)}\Big],\vspace{2mm}&
\end{align*}
where\begin{align*}
\Lambda=\dfrac{\lambda^2+\Delta}{2a\lambda}.
\end{align*}
\end{description}
In a similar fashion, we obtain
\begin{eqnarray*}
\lambda \dfrac{a}{b} T_{x+1}+\left(c-\dfrac{ad}{b}-\dfrac{\lambda^2}{b}\right)T_{x} +\lambda\dfrac{ d}{b}T_{x-1} =0\hspace{5mm}(x\geq n+2).
\end{eqnarray*} 
Then we have
\begin{description}
\item[(i)] $\lambda^2 \neq ad+bc\pm2\sqrt{abcd}$ case
\begin{align*}
T_{x}&=\dfrac{1}{\Lambda_{+} -\Lambda_{-}}\Big\{
\Lambda^{x-(n+1)}_{+} ( T_{n+2}-\Lambda_{-} T_{n+1} ) -\\
&\hspace{50mm}
\Lambda^{x-(n+1)}_{-} ( T_{n+2}-\Lambda_{+} T_{n+1} )\Big\}\hspace{5mm}(x\geq n+1),&
\end{align*}
where
\begin{align*}
\Lambda_{\pm}=\dfrac{\lambda^2+\Delta\pm\sqrt{(\lambda^2+\Delta)^2-4\lambda^{2}ad}}{2a\lambda}.
\end{align*}
\item[(ii)] $\lambda^2 = ad+bc\pm2\sqrt{abcd}$ case
\begin{align*}
T_x &= \Lambda^{x-(n+2)} \Big[T_{n+2}+\{x-(n+2)\}{(T_{n+2}-\Lambda T_{n+1})}\Big]\hspace{5mm}(x\geq n+1),&
\end{align*}
where\begin{align*}
\Lambda=\dfrac{\lambda^2+\Delta}{2a\lambda}.
\end{align*}
\end{description}


\section{Examples}\label{exam}
In this section, we give two examples. The first model is a QW with two defects. The second model is the Hadamard walk with three defects which is a generalization of the model proposed by Wojcic et al \cite{WojcikEtAl2004}.
\subsection{QW with two defects}

From now on, we consider the space-inhomogeneous QW whose quantum coin is determined by
\begin{align*}
U_x=
\begin{cases}
\begin{bmatrix}
\cos{\theta}&\sin{\theta}\\
\sin{\theta}&-\cos{\theta}
\end{bmatrix}&(x=-m,m),\\\\
\begin{bmatrix}
1&0\\
0&-1
\end{bmatrix}&(x\neq -m,m),
\end{cases}
\end{align*}
for $m\in\mathbb{Z}_>, \theta\neq\frac{\pi}{2}$. From Theorem \ref{thm:3.1} and Proposition \ref{pro:3.2}, we have
\begin{align*}
D^+ =
\begin{bmatrix}
\lambda&0\\
0&-\frac{1}{\lambda}
\end{bmatrix},\ 
D^{+}_{m}=
\begin{bmatrix}
\frac{\lambda}{\cos{\theta}}&\frac{\sin{\theta}}{\lambda \cos{\theta}}\\
0&-\frac{1}{\lambda}
\end{bmatrix},\ 
D^{+}_{m+1}=
\begin{bmatrix}
\lambda&0\\
\frac{\sin{\theta}}{\lambda}&-\frac{\cos{\theta}}{\lambda}
\end{bmatrix},\\\\
D^{-}_{-m}=
\begin{bmatrix}
\frac{1}{\lambda}&0\\
\frac{\sin{\theta}}{\lambda \cos{\theta}}&-\frac{\lambda}{\cos{\theta}}
\end{bmatrix},\ 
D^{-}_{-(m+1)}=
\begin{bmatrix}
\frac{\cos{\theta}}{\lambda}&\frac{\sin{\theta}}{\lambda}\\
0&-\lambda
\end{bmatrix},\ 
D^- =
\begin{bmatrix}
\frac{1}{\lambda}&0\\
0&-\lambda
\end{bmatrix}.
\end{align*}
By Theorem \ref{thm:3.1}, the amplitude becomes
\begin{eqnarray*}
\Psi(x)=
\begin{cases}
\displaystyle \prod_{y=1}^{x} D^{+}_{y} \Psi(0)&(x\geq 1),\\
\Psi(0)&(x=0),\\
\displaystyle \prod_{y=-1}^{x} D^{-}_{y} \Psi(0)&(x\leq -1),
\end{cases}
\end{eqnarray*}
i.e,
\begin{align*}
\begin{bmatrix}
\Psi^L (x)\\
\Psi^R (x)
\end{bmatrix}=
\begin{cases}
\begin{bmatrix}
\lambda^x \alpha\\
(-\frac{1}{\lambda})^x \beta
\end{bmatrix}&(0\leq x \leq m-1),\\\\
\begin{bmatrix}
\frac{1}{\cos{\theta}} \left\{ \lambda^m \alpha-\sin\theta(-\frac{1}{\lambda})^m\beta \right\}\\
(-\frac{1}{\lambda})^m\beta
\end{bmatrix}&(x=m),\\\\
\begin{bmatrix}
\frac{\lambda}{\cos\theta} \left\{ \lambda^m \alpha - \sin\theta (-\frac{1}{\lambda})^m \beta \right\}\\
\frac{1}{\lambda\cos\theta} \left\{ \sin\theta \lambda^m \alpha - (-\frac{1}{\lambda})^m \beta \right\}
\end{bmatrix}&(x=m+1),\\\\
\begin{bmatrix}
\lambda^{x-(m+1)} \Psi^L (m+1)\\
(-\frac{1}{\lambda})^{x-(m+1)} \Psi^R (m+1)
\end{bmatrix}&(x\geq m+2),
\end{cases}
\end{align*}

\begin{align*}
\begin{bmatrix}
\Psi^L (x)\\
\Psi^R (x)
\end{bmatrix}=
\begin{cases}
\begin{bmatrix}
(\frac{1}{\lambda})^{-x}\alpha\\
(-\lambda)^{-x}\beta
\end{bmatrix}&(-m+1\leq x \leq 0),\\\\
\begin{bmatrix}
\frac{1}{\lambda^m} \alpha\\
\frac{1}{\cos\theta} \left\{ \sin\theta \frac{1}{\lambda^m}\alpha + (-\lambda)^m\beta \right\}
\end{bmatrix}&(x=-m),\\\\
\begin{bmatrix}
\frac{1}{\lambda\cos\theta} \left\{ \frac{1}{\lambda^m}\alpha+\sin\theta (-\lambda)^m \beta \right\}\\
-\frac{\lambda}{\cos\theta} \left\{ \sin\theta \frac{1}{\lambda^m}\alpha +(-\lambda)^m \beta \right\}
\end{bmatrix}&(x=-m-1),\\\\
\begin{bmatrix}
(\frac{1}{\lambda})^{-x-(m+1)}\Psi^L (-(m+1))\\
(-\lambda)^{-x-(m+1)}\Psi^R (-(m+1))
\end{bmatrix}&(x\leq-m-2).
\end{cases}
\end{align*}
Furthermore, a stationary measure $\mu$ is given by
\begin{align*}
\mu(x)=\phi(\Psi)(x)=||\Psi(x)||^2\ \ \ (x\in\mathbb{Z}).
\end{align*}
For example, if $\alpha=1/\sqrt{2}, \beta=i/\sqrt{2}$ and $\theta=\pi/4$, we get the stationary measure $\mu$ given by
\begin{align*}
\mu(x)=
\begin{cases}
1&(x\in\mathbb{Z}_{[-(m-1),m-1]}),\\
2&(x=\pm m),\\
3&(x\notin\mathbb{Z}_{[-m,m]}).
\end{cases}
\end{align*} 
This stationary measure is not uniform measure.

\subsection{Hadamard walk with three defects}

Next, we consider the space-inhomogeneous QW whose quantum coin is determined by
\begin{align*}
U_x=
\begin{cases}
\omega H&(x\in\mathbb{Z}_{[-1,1]}),\vspace{3mm}\\
H&(x\notin\mathbb{Z}_{[-1,1]}),
\end{cases}
\end{align*}
with $\omega=e^{2i\pi \phi}\ (\phi\in[0,1))$, where
\begin{align*}
H=\frac{1}{\sqrt{2}}
\begin{bmatrix}
1&1\\
1&-1
\end{bmatrix}.
\end{align*}
 In particular, if $\phi=0\ (i.e.,\ \omega=1)$, then this space-homogeneous QW is called the Hadamard walk.
Endo and Konno \cite{ek2014} investigated the stationary measures of Hadamard walk with one defect introduced by Wojcik et al. \cite{WojcikEtAl2004} via a different approach, i.e., the splitted generating function method.\\
 From Theorem \ref{thm:3.1} and Proposition \ref{pro:3.2}, we have
\begin{align*}
D^+=
\begin{bmatrix}
\frac{2\lambda^2-1}{\sqrt{2}\lambda}&\frac{1}{\sqrt{2}\lambda}\\
\frac{1}{\sqrt{2}\lambda}&-\frac{1}{\sqrt{2}\lambda}
\end{bmatrix},\ 
D^{+}_2 =
\begin{bmatrix}
\frac{2\lambda^2-\omega}{\sqrt{2}\lambda}&\frac{\omega}{\sqrt{2}\lambda}\\
\frac{\omega}{\sqrt{2}\lambda}&-\frac{\omega}{\sqrt{2}\lambda}
\end{bmatrix},\ 
D^{+}_1 =
\begin{bmatrix}
\frac{2\lambda^2-\omega^2}{\sqrt{2}\lambda}&\frac{\omega}{\sqrt{2}\lambda}\\
\frac{\omega}{\sqrt{2}\lambda}&-\frac{\omega}{\sqrt{2}\lambda}
\end{bmatrix},\\
D^{-}_{-1}=
\begin{bmatrix}
\frac{\omega}{\sqrt{2}\lambda}&\frac{\omega}{\sqrt{2}\lambda}\\
\frac{\omega}{\sqrt{2}\lambda}&-\frac{2\lambda^2-\omega^2}{\sqrt{2}\omega\lambda}
\end{bmatrix},\ 
D^{-}_{-2}=
\begin{bmatrix}
\frac{\omega}{\sqrt{2}\lambda}&\frac{\omega}{\sqrt{2}\lambda}\\
\frac{\omega}{\sqrt{2}\lambda}&-\frac{2\lambda^2-\omega}{\sqrt{2}\omega\lambda}
\end{bmatrix},\ 
D^{-}=
\begin{bmatrix}
\frac{1}{\sqrt{2}\lambda}&\frac{1}{\sqrt{2}\lambda}\\
\frac{1}{\sqrt{2}\lambda}&-\frac{2\lambda^2-1}{\sqrt{2}\lambda}
\end{bmatrix}.
\end{align*}
Furthermore, the amplitude is given by
\begin{align*}
\Psi(x)=
\begin{cases}
\bigg\{ (D^{+}) ^{x-2}\bigg\}\displaystyle \prod_{y=1}^{2}D^{+}_{y} \Psi(0)&(3\leq x),\\
\displaystyle \prod_{y=1}^{x}D^{+}_{y} \Psi(0)&(1\leq x \leq 2),\\
\Psi(0)&(x=0),\\
\displaystyle \prod_{y=-1}^{x}D^{-}_{y} \Psi(0)&(-2 \leq x \leq -1),\\
\bigg\{ (D^{-})^{-x-2} \bigg\}\displaystyle \prod_{y=-1}^{-2}D^{-}_{y} \Psi(0)&(x\leq -3),
\end{cases}
\end{align*}
Next, by Proposition \ref{pro:3.3}, we consider the amplitude of $3\leq x$ and $x\leq-3$. If characteristic equation of Eq.(\ref{ukiyoe7}) has a multiple root, then 
$
\lambda=
e^{\pm \frac{\pi i}{4}}, -e^{\pm \frac{\pi i}{4}}. \label{tokyo1}
$
Therefore the amplitude is given by

\begin{description}
\item[(i)]$\lambda \neq
e^{\pm \frac{\pi i}{4}}, -e^{\pm \frac{\pi i}{4}}$ (distinct root) case\\
\begin{align*}
&\begin{bmatrix}
\Psi^{L}(x)\\
\Psi^{R}(x)
\end{bmatrix}\\
&=
\begin{cases}
A
\begin{bmatrix}
\Lambda^{x-1}_{+} (\Psi^{L}(2)-\Lambda_{-}\Psi^{L}(1))-
\Lambda^{x-1}_{-}(\Psi^{L}(2)-\Lambda_{+}\Psi^{L}(1))\vspace{3mm}\\
\Lambda^{x-2}_{+} (\Psi^{R}(3)-\Lambda_{-}\Psi^{R}(2))-
\Lambda^{x-2}_{-}(\Psi^{R}(3)-\Lambda_{+}\Psi^{R}(2))
\end{bmatrix}\ \ (x\geq3),\\\\
A
\begin{bmatrix}
\Gamma^{-x-2}_{+} (\Psi^{L}(-3)-\Gamma_{-}\Psi^{L}(-2))-
\Gamma^{-x-2}_{-}(\Psi^{L}(-3)-\Gamma_{+}\Psi^{L}(-2))\vspace{3mm}\\
\Gamma^{-x-1}_{+} (\Psi^{R}(-2)-\Gamma_{-}\Psi^{R}(-1))-
\Gamma^{-x-1}_{-}(\Psi^{R}(-2)-\Gamma_{+}\Psi^{R}(-1))
\end{bmatrix}\\\hspace{90mm}(x\leq-3),
\end{cases}
\end{align*}\\
\item[(ii)]$\lambda=
e^{\pm \frac{\pi i}{4}}, -e^{\pm \frac{\pi i}{4}}$ (multiple root) case.
\begin{align*}
&\begin{bmatrix}
\Psi^{L}(x)\\
\Psi^{R}(x)
\end{bmatrix}\\
&=
\begin{cases}
\begin{bmatrix}
B^{x-2}\left\{ \Psi^L (2)+(x-2)(\Psi^L (2)-B\Psi^L (1)) \right\}\vspace{3mm}\\
B^{x-3}\left\{ \Psi^R (3)+(x-3)(\Psi^R (3)-B\Psi^R (2)) \right\}
\end{bmatrix}&(x\geq3),\\\\
\begin{bmatrix}
(-B)^{-x-3}\left\{ \Psi^L (-3)+(-x-3)(\Psi^L (-3)+B\Psi^L (-2)) \right\}\vspace{3mm}\\
(-B)^{-x-2}\left\{ \Psi^R (-2)+(-x-2)(\Psi^R (-2)+B\Psi^R (-1)) \right\}
\end{bmatrix}&(x\leq-3),
\end{cases}
\end{align*}
\end{description}
where
\begin{align*}
\Lambda_{\pm}=&\frac{\lambda^2-1\pm\sqrt{\lambda^4+1}}{\sqrt{2}\lambda},\ \Gamma_{\pm}=-\Lambda_{\mp},\ A=\dfrac{\lambda}{\sqrt{2(\lambda^4+1)}},\ 
B=\frac{\lambda^2-1}{\sqrt{2}\lambda},\\
\begin{bmatrix}
\Psi^{L}(1)\\
\Psi^{R}(1)
\end{bmatrix}=&
\begin{bmatrix}
\frac{2 \alpha {{\lambda}^{2}}+\beta {{\omega}^{2}}-\alpha {{\omega}^{2}}}{\sqrt{2} \omega b \lambda}\\\\
 -\frac{\left( \beta-\alpha\right)  \omega}{\sqrt{2} \lambda}
\end{bmatrix},\ 
\begin{bmatrix}
\Psi^L (2)\\
\Psi^R (2)
\end{bmatrix}=
\begin{bmatrix}
\frac{2\alpha \lambda^4-\left\{ (\alpha-\beta)\omega^2+\alpha\omega \right\}\lambda^2+(\alpha-\beta)\omega^3}{\omega\lambda^2}\\\\
\frac{\alpha\lambda^2-(\alpha-\beta)\omega^2}{\lambda^2}
\end{bmatrix},\\
\begin{bmatrix}
\Psi^L (-1)\\
\Psi^R (-1)
\end{bmatrix}=&
\begin{bmatrix}
\frac{(\alpha+\beta)\omega}{\sqrt{2}\lambda}\\\\
-\frac{2\beta\lambda^2-\beta\omega^2-\alpha\omega^2}{\sqrt{2}\omega\lambda}
\end{bmatrix},\ 
\begin{bmatrix}
\Psi^L (-2)\\
\Psi^R (-2)
\end{bmatrix}=
\begin{bmatrix}
-\frac{\beta\lambda^2-(\alpha+\beta)\omega^2}{\lambda^2}\\\\
\frac{2\beta\lambda^4-\left\{ (\alpha+\beta)+\beta\omega \right\}\lambda^2+(\alpha+\beta)\omega^3}{\omega\lambda^2}
\end{bmatrix},\\
\end{align*}
and
\begin{align*}
\Psi^R (3)=&
\frac{2\alpha\lambda^4 - \left\{ (\alpha-\beta)\omega^2+2\alpha\omega \right\}\lambda^2 +2(\alpha-\beta)\omega^3 }{\sqrt{2}\omega\lambda^3},\\
\Psi^L (-3)=&
\frac{2\beta\lambda^4-\left\{ (\alpha+\beta)\omega^2+2\beta\omega \right\}\lambda^2+2(\alpha+\beta)\omega^3}{\sqrt{2}\omega\lambda^3}.
\end{align*}
Furthermore, a stationary measure $\mu$ is given by
\begin{align*}
\mu(x)=\phi(\Psi)(x)=||\Psi(x)||^2\ \ \ (x\in\mathbb{Z}).
\end{align*}

\section{Summary}\label{summ}
In this paper, we obtained stationary measures for the two-state space-inhomogeneous QWs on $\ZM$ by solving the corresponding eigenvalue problem (Theorem 3.1). From this result follow several interesting corollaries. For example, we got  a stationary measure $\mu\in\mathcal{M}_s(U^{(s)})\cap\mathcal{M}_{et}(U^{(s)})$ in Corollary \ref{cor:3.4} (i). This case is a generalization of the model studied by Konno et al.\cite{KLS2013}. On the other hand, we obtained a stationary measure  $\mu\in\mathcal{M}_s(U^{(s)})\cap\mathcal{M}_{qpt}(U^{(s)})$ in Corollary \ref{cor:3.4} (ii). This case is a generalization of the model considered by Konno and Takei \cite{kt2015}.

As a future work, it would be fascinating  to investigate the stationary measure of the multi-state space-inhomogeneous QW on general graphs.

\begin{small}
\bibliographystyle{jplain}

\end{small}

\end{document}